\documentclass{emulateapj}\usepackage{apjfonts}

\newcommand\kms{km~s$^{-1}$}

\newcommand\etal{{et~al.}} 
\newcommand\mM{\ifmmode(m{-}M)\else$(m{-}M)$\fi}

\newcommand\hst{{\it HST}}
\newcommand\zacs{\ifmmode z_{850}\else$z_{850}$\fi}
\newcommand\Iacs{\ifmmode I_{814}\else$I_{814}$\fi}
\newcommand\iacs{\ifmmode i_{775}\else$i_{775}$\fi}
\newcommand\gacs{\ifmmode g_{475}\else$g_{475}$\fi}
\newcommand\racs{\ifmmode r_{625}\else$r_{625}$\fi}
\newcommand\vacs{\ifmmode V_{606}\else$V_{606}$\fi}
\newcommand\gz{{\ifmmode{(g_{475}{-}z_{850})}\else$(g_{475}{-}z_{850})$\fi}}
\newcommand\gzacs{\gz}
\newcommand\riacs{{\ifmmode{r_{625}{-}i_{775}}\else$r_{625}{-}i_{775}$\fi}}
\newcommand\rzacs{{\ifmmode{r_{625}{-}z_{850}}\else$r_{625}{-}z_{850}$\fi}}
\newcommand\izacs{{\ifmmode{i_{775}{-}z_{850}}\else$i_{775}{-}z_{850}$\fi}}
\newcommand\vzacs{{\ifmmode{V_{606}{-}z_{850}}\else$V_{606}{-}z_{850}$\fi}}
\newcommand\vi{{\ifmmode{(V{-}I)}\else$(V{-}I)$\fi}}
\newcommand\gI{{\ifmmode{(g{-}I)}\else$(g{-}I)$\fi}}
\newcommand\gIacs{{\ifmmode{(g_{475}{-}I_{814})}\else$(g_{475}{-}I_{814})$\fi}}
\newcommand\mbar{\ensuremath{\overline{m}}}

\newcommand\mbarz{\ensuremath{\overline{m}_z}}
\newcommand\mbarIacs{\ensuremath{\overline{m}_{814}}}
\newcommand\mbarzacs{\ensuremath{\overline{m}_{850}}}
\newcommand\Mbar{\ensuremath{\overline{M}}}
\newcommand\MbarI{\ensuremath{\overline{M}_I}}

\newcommand\MbarIacs{\ensuremath{\overline{M}_{814}}}
\newcommand\Mbarzacs{\ensuremath{\overline{M}_{850}}}
\newcommand\lta{\lesssim}

\newcommand\cote{C{\^ o}t{\' e}}
\newcommand\jordan{Jord{\'a}n}

\def\fcsv{\hbox{ACSFCS-V}}

\shortauthors{{Blakeslee et al.}}
\shorttitle{SBF Distances}
\slugcomment{Accepted for publication in ApJ}

\begin{document}

\title{Surface Brightness Fluctuations in the Hubble Space Telescope
  ACS/WFC F814W Bandpass \\
and an Update on Galaxy~Distances\altaffilmark{*}}

\altaffiltext{*}{Based on observations with the NASA/ESA \textit{Hubble Space
  Telescope}, obtained from the Space Telescope Science
  Institute, which is operated by AURA, Inc.,
  under NASA contract NAS 5-26555.}

\author{John P.~Blakeslee\altaffilmark{1,2},
Michele~Cantiello\altaffilmark{3},
Simona~Mei\altaffilmark{4,5},
Patrick~C\^ot\'e\altaffilmark{1},
Regina Barber~DeGraaff\altaffilmark{2},
Laura~Ferrarese\altaffilmark{1},
Andr\'es~Jord\'an\altaffilmark{6},
Eric~W.~Peng\altaffilmark{7},
John~L.~Tonry\altaffilmark{8},
Guy~Worthey\altaffilmark{2}
}
\altaffiltext{1}{Herzberg Institute of Astrophysics, National Research Council of Canada, Victoria, BC V9E\,2E7, Canada; John.Blakeslee@nrc.ca}
\altaffiltext{2}{Department of Physics and Astronomy, Washington State University, Pullman, WA 99163-2814}
\altaffiltext{3}{INAF-Osservatorio Astronomico di Teramo, via M. Maggini, I-64100, Teramo, Italy}
\altaffiltext{4}{University of Paris Denis Diderot, 75205 Paris Cedex 13, France}
\altaffiltext{5}{GEPI, Observatoire de Paris, Section de Meudon, 5 Place J.\ Janssen, 92195 Meudon Cedex, France}
\altaffiltext{6}{Departamento de Astronom{\'i}a y Astrof{\'i}sica, Pontificia Universidad Cat{\'o}lica de Chile, Santiago 22, Chile}
\altaffiltext{7}{Department of Astronomy, Peking University, Beijing 100871, China}
\altaffiltext{8}{Institute for Astronomy, University of Hawaii, Honolulu, HI 96822}

\begin{abstract}
We measure surface brightness fluctuation (SBF) magnitudes in the F814W filter
and \gIacs\ colors for nine bright early-type Fornax cluster galaxies imaged
with the \textit{Hubble Space Telescope} Advanced Camera for Surveys (ACS).
The goal is to achieve the first systematic SBF calibration for the ACS/F814W
bandpass. Because of its much higher throughput, F814W is more efficient for
SBF studies of distant galaxies than the ACS/F850LP bandpass that has been used
to study nearby systems.
Over the color range spanned by the sample galaxies,
$1.06<\gIacs<1.32$ (AB mag), the dependence of
SBF magnitude \mbarIacs\ on \gIacs\ is linear to a good
approximation, with slope $\sim2$.  When the F850LP SBF distance measurements
from the ACS Fornax Cluster Survey are used to derive absolute \MbarIacs\
magnitudes, the dependence on \gIacs\ becomes extremely tight, with a
slope of $1.8\pm0.2$ and scatter of 0.03~mag.  The small
observed scatter indicates both that the estimated random errors are correct,
and that the intrinsic deviations from the SBF--color relation are strongly
correlated between the F814W and F850LP bandpasses, as expected.  
The agreement with predictions from stellar population models is good, both in slope 
and zero point, indicating that our mean Fornax distance of 20~Mpc is accurate.  
The models predict curvature in the relation beyond the color limits of our
sample; thus, the linear calibration should not be extrapolated naively.
In the Appendices, we reconsider the Tonry ground-based and Jensen NICMOS SBF
distance catalogues; we provide a correction formula to ameliorate the small
apparent bias in the former, and the offset needed to make the latter
consistent with other SBF studies.
We also tabulate two new SBF distances to galaxies observed in the ACS Virgo
Cluster Survey.
\end{abstract}


\section{Introduction}
\label{sec:intro}

The measurement of galaxy distances has been a central problem in
astronomy ever since the discovery of ``spiral nebulae'' by Lord Rosse in the
mid-nineteenth century.  Because geometric methods were of no avail, the
nature of these objects remained a matter of conjecture for many decades. 
Only with the work of Henrietta Swan Leavitt did accurate photometric
distance estimates to resolved stellar systems become a reality.  
Her painstaking photographic measurements of the light curves 
for 1800 variable stars (and hundreds of comparison stars)
in the Magellanic Clouds led her to note that for variables of a certain
class, the brighter ones had longer temporal periods (Leavitt 1908). 
The Leavitt Law, relating the luminosities of Cepheid variables to their
periods,
proved to be the key to the mystery of the spiral nebulae, locating them firmly beyond
the spatial extent of our own Galaxy; thus it was also the key to comprehending the true
scale of the Universe, eventually accomplished with the work of Hubble~(1926,\,1929).

Accurate galaxy distances remain of fundamental importance to most
astrophysical applications.  Errors in distance estimates generally translate
into equal or larger fractional uncertainties in derived quantities such as
masses (of everything from central black holes to dark matter haloes), linear
sizes, dynamical timescales, star formation rates, and ages.  Thus, accurate
distance estimation is essential for inter-comparing the physical properties of
galaxies in the local volume where redshift is not a reliable indicator of
distance, and for comparing the properties of nearby galaxies to those at high
redshift.  Of course, they are also essential for mapping local
three-dimensional structures and velocity fields.

There are a variety of distance indicators that can be used within and beyond
the limit of classical Cepheid distances; Freedman \& Madore (2010) provide a recent 
overview of the methods.  One of the most accurate of these is surface brightness
fluctuations (SBF; Tonry \& Schneider 1988; for a historical review, see
Blakeslee \etal\ 2009, hereafter \fcsv).
The SBF method measures the intrinsic variance in a galaxy
image resulting from the random variations in the numbers and luminosities of
the stars falling within individual pixels of the image.
This variance is normalized by the local galaxy surface brightness and then
converted to the apparent SBF magnitude~\mbar.  The distance modulus,
\mbar$\,{-}$\Mbar\ follows once the absolute magnitude \Mbar\ is known.

The value of \Mbar\ in a given bandpass depends on the stellar
population properties.  For early-type systems, a single broad-baseline color
generally suffices for characterizing the stellar population (e.g., Tonry
\etal\ 1997; Ajhar \etal\ 1997; Ferrarese \etal\ 2000; Blakeslee \etal\ 2001; 
Mei \etal\ 2005b; \fcsv).  The zero-point of the $I$-band \Mbar\ calibration has been
tied directly to the Cepheid distance scale to an accuracy of 0.08 mag, or
$\sim\,$4\% in distance, from ground-based SBF measurements in spiral bulges
(Tonry et al. 2000).
For distance estimation purposes, SBF is best measured in the reddest optical 
bandpasses.  The fluctuations arise
mainly from red giant branch stars and thus are brighter at redder
wavelengths.  However, the stellar population dependence becomes 
more complicated in the near-infrared (Jensen \etal\ 2003), which adds 
intrinsic scatter to that application of the method.  Thus, the red end of
the optical spectrum is something of a ``sweet spot'' for SBF.

The SBF method has been an important part of two recent 
\textit{Hubble Space Telescope} (\hst) surveys of early-type
cluster galaxies carried out with the Advanced Camera for Surveys' Wide Field Channel (ACS/WFC).
The 100-orbit ACS Virgo Cluster
Survey (ACSVCS; \cote\ \etal\ 2004) and the 43-orbit ACS Fornax Cluster Survey
(ACSFCS; Jord\'an \etal\ 2007) both observed one galaxy per
orbit, with the time split between the F475W and F850LP filters.
The SBF analysis was performed on the F850LP ($z_{850}$-band) images, 
while the second bandpass enabled the calibration of the SBF $z_{850}$
magnitude in terms of \gz\ color (Mei \etal\ 2005b; \fcsv).
These single-orbit observations were adequate at the distances of
the Virgo and Fornax clusters,
but for galaxies significantly farther away, the relatively low throughput 
of F850LP can make the required exposure times prohibitive.
The F814W filter then becomes more attractive for SBF work
simply because the higher total throughput of this passband reduces the
exposure time by at least a factor of two.  
However, although there have been some ACS/WFC measurements of SBF in this bandpass
(Cantiello \etal\ 2007a,b;
Barber DeGraaff \etal\ 2007; Biscardi \etal\ 2008), there has been no systematic empirical
calibration of the SBF method in F814W similar to that in F850LP.
The main goal of the present work is to provide such a calibration that will be
applicable to the bright early-type galaxies targeted at large distances.

The following section presents our observational sample and basic imaging
reductions.  Section~\ref{sec:sbf} describes our SBF and
color measurements in detail and compares the results to our
previous F850LP measurements.
Section~\ref{sec:models} provides a comparison with predictions from stellar
population models and discusses the uncertainty in the absolute distance to
the Fornax cluster.
Section~\ref{sec:summary} presents the conclusions from this study.  In
Appendix~\ref{app:a}, we revisit the ground-based SBF distances from Tonry
\etal\ (2001) and provide a simple correction formula for possible bias in
those distances.   Finally, Appendix~\ref{app:b} presents two additional SBF
distance measurements from the ACSVCS F475W and F850LP imaging~data.

\section{Observations and Image Reductions}
\label{sec:obs}

The proximity ($\sim\,$20 Mpc) and compactness of the elliptical-rich Fornax
cluster makes it the ideal target for calibrating the stellar population
dependence of the SBF method in any bandpass (e.g., Tonry 1991; \fcsv).
As we are mainly interested in the calibration for luminous early-type galaxies,
we selected the eight brightest Fornax galaxies as listed by \jordan\ \etal\
(2007), all of which have $M_B<-19$.  Because of the precision of ACS/WFC
SBF measurements, this sample was deemed suitable for deriving a calibration
accurate to about $\pm0.02$ mag over the normal color range for giant
ellipticals. 

As part of \hst\ program GO-10911, we observed six fields with the ACS/WFC
targeting large galaxies in the Fornax cluster.  Each observation consisted of
three dithered exposures in the F814W filter totaling 1224\,s and two
dithered exposures in F475W totaling 680\,s.  The galaxy NGC\,1375 is
separated by only $2\farcm4$ from its neighbor NGC\,1374, and
we were able to fit them both along the 4\farcm7 diagonal of a single ACS/WFC
field of view.  NGC\,1374 was placed in the upper left quadrant of the field, on the
same CCD chip on which the other giant galaxies were centered, while NGC\,1375
was simultaneously in the lower right quadrant.
Although NGC\,1375 is 1.4~mag fainter than the limit of the other
eight, it provides extra leverage for constraining the calibration slope and
was available gratis. 
In addition, we downloaded suitable
F814W data from the \hst\ archive on the Fornax galaxies
NGC\,1316 (from GO-9409, PI: Goudfrooij) and 
NGC\,1344 (from GO-9399, PI: Carter) and suitable F475W data on
the same two galaxies from GO-10217 (PI: \jordan).
Table~\ref{tab:tab_obs} summarizes the properties and
basic observational details for our nine target galaxies, including the
\zacs\ SBF distances from the \fcsv.

The exposures were bias and dark current subtracted and flatfielded by the
standard STScI pipeline processing.
The calibrated ``flt'' data were then run through the Apsis pipeline (Blakeslee
\etal\ 2003) to produce geometrically rectified, clean, combined images.
Apsis uses Drizzle (Fruchter \& Hook 2002) for the geometric transformation of
the images, and we use the Lanczos3 interpolation kernel in order to preserve
the essential properties of the image power spectra (see Mei \etal\ 2005a).
All photometry reported here is on the AB system, calibrated based on Sirianni
et al.\ (2005) and Bohlin (2007) with the revised zero points from
the ACS instrument
website\footnote{http://www.stsci.edu/hst/acs/analysis/zeropoints} for data
obtained before July~2006.  In particular, we adopted: 
$m_1(\hbox{F475W}) = 26.0809$, $m_1(\hbox{F814W}) = 25.9593$,
where $m = -2.5\,\log(f) + m_1$, and $f$ is in
electrons per second.
The ACS/WFC F814W and F475W bandpasses closely approximate the Kron-Cousins
$I$ and Sloan Digital Sky Survey $g$ bands, so we
refer to magnitudes in these bandpasses as \gacs\ and \Iacs, respectively.
We corrected for Galactic extinction using the dust maps from Schlegel \etal\
(1998) with the extinction ratios given by Sirianni \etal\ (2005).
The extinction in this direction is small, $E(B{-}V)\lta0.02$~mag.

\section{SBF Analysis}
\label{sec:sbf}

\subsection{Measurements within Annuli}
\label{ssec:annuli}

There have been a variety of implementations of the SBF analysis within 
different reduction packages (e.g., Tonry \etal\ 1990; Lorenz \etal\ 1993;
Pahre \etal\ 1999; Nielsen \& Tsvetanov 2000; Liu \etal\ 2002; Mieske \etal\
2003; Mei \etal\ 2005a; Cantiello \etal\ 2005; Dunn \& Jerjen 2006), but 
few detailed comparisons between them.  One notable exception is the
comparison made by Pahre \etal\ (1999) of their SBF results for one galaxy
observed with HST/WFPC2 with the results for the same observation using the
Tonry \etal\ (1990, 1997) software.  They found good agreement for this one
case.  Otherwise, the comparisons have mainly been between sets of published
numbers, rather than tests of different analysis procedures applied to the
same data.

\begin{figure*}\epsscale{0.95}
\plotone{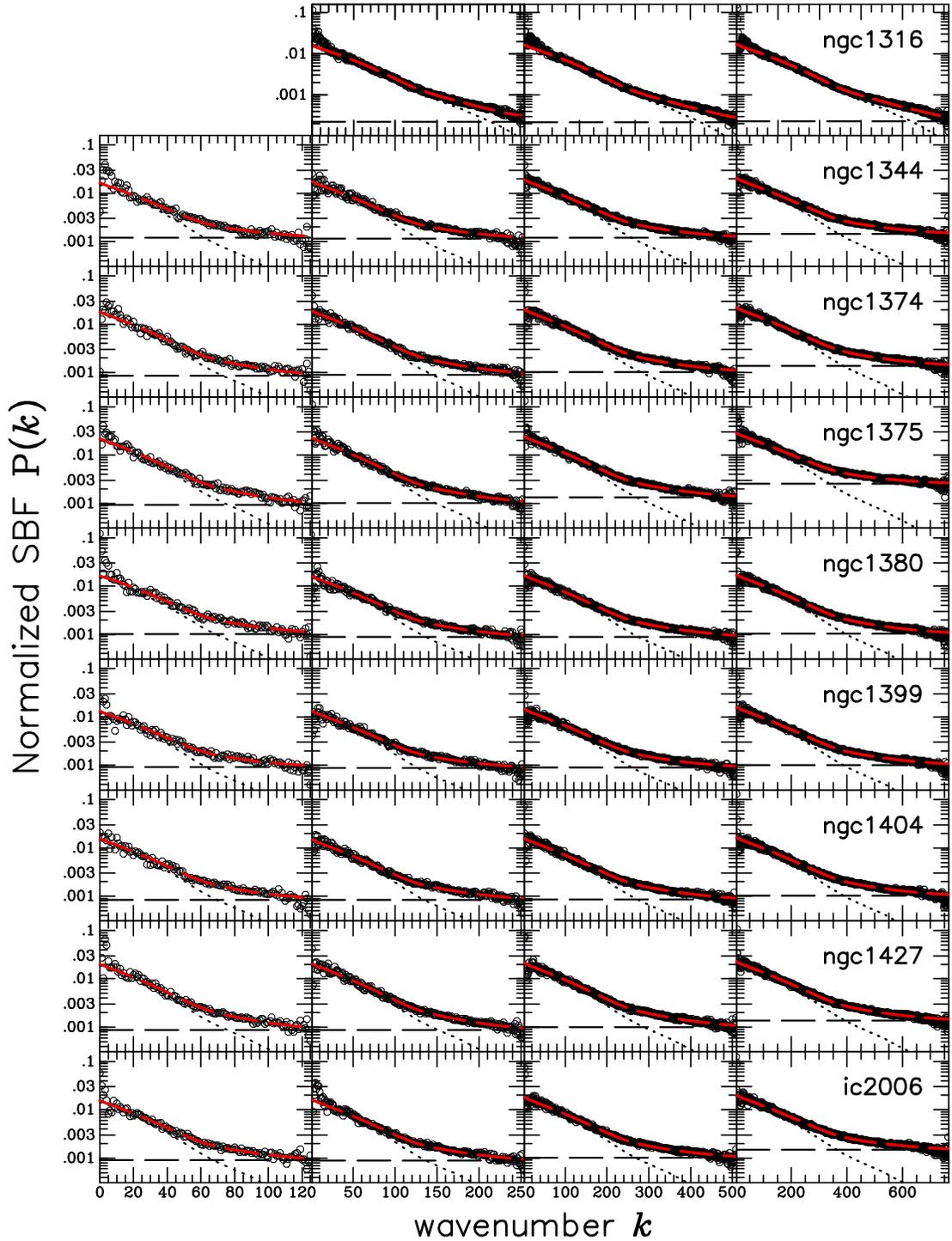}
\caption{\small
Power spectrum measurements for different annuli within our program galaxies (labeled);
the radii of the annuli increase from left to right, as described in the text.
The red curves show our fits to the power spectra.  Because of the significant
dust features, the annuli for NGC\,1316 differ from the other galaxies', and
only three were used.  The power spectra are normalized by the local galaxy
surface brightness, so that the amplitude $P_0$ of the PSF component (dotted
curve) corresponds to the SBF magnitude \mbar, after a small $\lta2$\%
correction for background variance.
\label{fig:powspec}}
\vspace{1cm}
\end{figure*}

In order to ensure that our calibration is as general as possible,
we performed two complete sets of SBF reductions on all the galaxies.  We first used 
the custom SBF analysis software developed by
J.~Tonry within the Vista
environment\footnote{http://www.ifa.hawaii.edu/$\sim$jt/soft.html} for the
ground-based SBF survey (Tonry \etal\ 1997) and applied most recently to
\hst/ACS data by Barber DeGraaff \etal\ (2007).
We refer to this set of reductions as the ``Tonry'' SBF analysis; details
can be found in the referenced works and several others that have used the same software
(e.g., Blakeslee \etal\ 1997; Jensen \etal\ 1998).  These reductions tend to 
be interactive, with the user selecting radial limits for the
analysis regions within each galaxy and power spectrum fitting limits.  Second, we
applied the identical SBF analysis code as used in \fcsv, which was only slightly
modified from the analysis routines developed and used by Mei \etal\
(2005a,b).  These routines are written in IDL and the SBF analysis is more
automated, which facilitates multiple runs to test for systematics effects.
As an example, Mei \etal\ (2005a) reran the analysis numerous times on
simulated images processed with Apsis to assess the effects of different
Drizzle interpolation kernels.

While there are no fundamental differences between the two analysis codes
likely to cause any significant disagreements, the current data set affords
a useful test of the level consistency that can be expected.
For both sets of SBF measurements, 
we model and subtract the galaxy, mask all visible sources, and
estimate the residual variance from undetected sources as detailed in our previous
works (Mei \etal\ 2005b; Barber DeGraaff \etal\ 2007; \jordan\ \etal\ 2004, 2007;
\fcsv).  These steps are the same for nearly all implementations of the SBF
method, and we used the same galaxy models and background variance corrections
for both sets of reductions.
For this sample of high-resolution, high-throughput F814W ACS imaging
data, the estimated background corrections ranged from 0.011 to
0.019~mag.  As in our previous works, we assign 25\% uncertainty to
these values, meaning that the contribution to the total error in the
SBF magnitude is less than 0.005~mag.  This contrasts strongly with
ground-based analyses, for which the source detection limits tend to be
much brighter and the background variance correction is often a major
source of uncertainty.
We masked dusty regions in the
galaxy images as described in Ferrarese \etal\ (2006) by taking the ratio of
the images in the two bands and identifying areas that significantly deviated
from the smooth color profile.  The dust masks were augmented by hand
as needed.  The two galaxies in our sample with significant dust features are
NGC\,1316 and NGC\,1380.  Thus, this comparison of SBF analyses does not
address possible problems with galaxy modeling or residuals due to dust or
other contaminants.  However, the consistency of the results as a function of
radius within a galaxy can be used to check for such problems.

Both the Tonry and IDL analysis procedures are performed in a series of
annuli with increasing radii. 
The inner radius of the innermost annulus was 6\farcs4 (128 pix) for all
galaxies except the dusty post-merger galaxy NGC\,1316, for which it 
was 12\farcs8.  The outer radius of each annulus was typically 
twice its inner radius,
although for the IDL code we used a maximum annular width of 256 pix to be more
similar to our past analyses, and only annuli where the galaxy \Iacs\ surface
brightness is at least twice the sky background were used.  For the Tonry code
analysis, we set the maximum outer radius somewhat arbitrarily at 1000~pix,
except in the case of the faintest galaxy NGC\,1375, for which it was 512~pix.
The two different analysis procedures independently calculate the
power spectra and fit the amplitude of the component that has been
convolved with the point spread function (PSF), which corresponds to the
SBF signal after correcting for the background variance.  
Section~\ref{ssec:psf} discusses systematic effects from different PSFs.

In fitting the power spectra, the IDL code omits the lowest wavenumbers,
those representing spatial scales larger than 20~pix, since these are
affected by the galaxy and large-scale background light subtraction, and
it also omits the several highest wavenumbers because of the pixel
correlation introduced by the geometric correction (see Mei \etal\ 2005
for details).  However, with the adopted Lanczos3 kernel, the pixel
correlation is very mild and the power spectrum fit is not very
sensitive to the maximum wavenumber, assuming that the PSF model is
accurate.  The Tonry code also omits the lowest numbers, although this
is done interactively by selecting the point where the fitted amplitude
$P_0$ of the PSF component becomes stable.  There is no option in this
code to omit the highest wavenumbers, so this is another difference in
the analyses.
Figure~\ref{fig:powspec} displays power spectra fits for multiple annuli within
the program galaxies from the IDL analysis code.  The signal is very strong and
well fitted in all cases; note the logarithmic scale.

\begin{figure}\epsscale{1.0}
\plotone{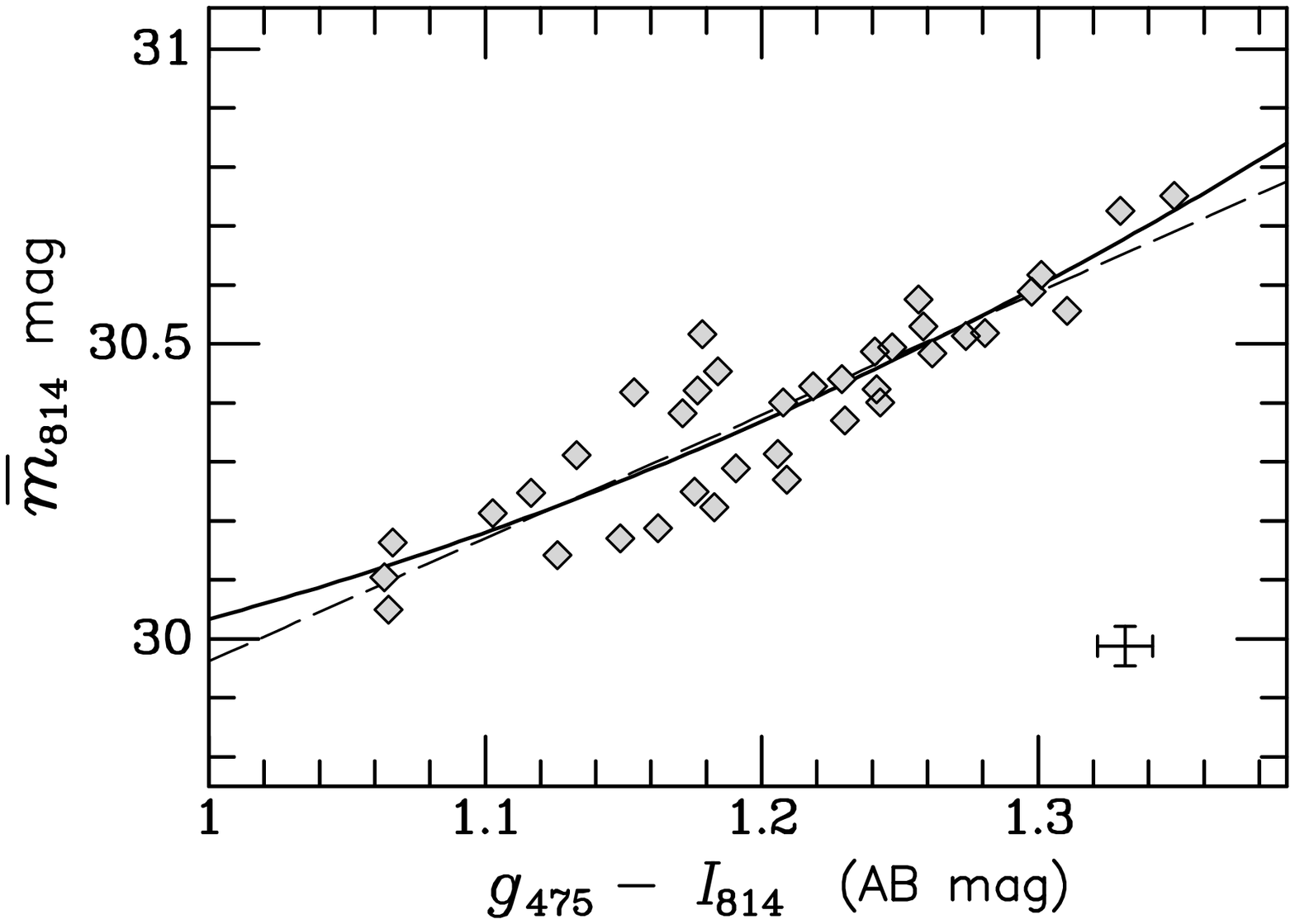}
\vspace{0.5cm}
\plotone{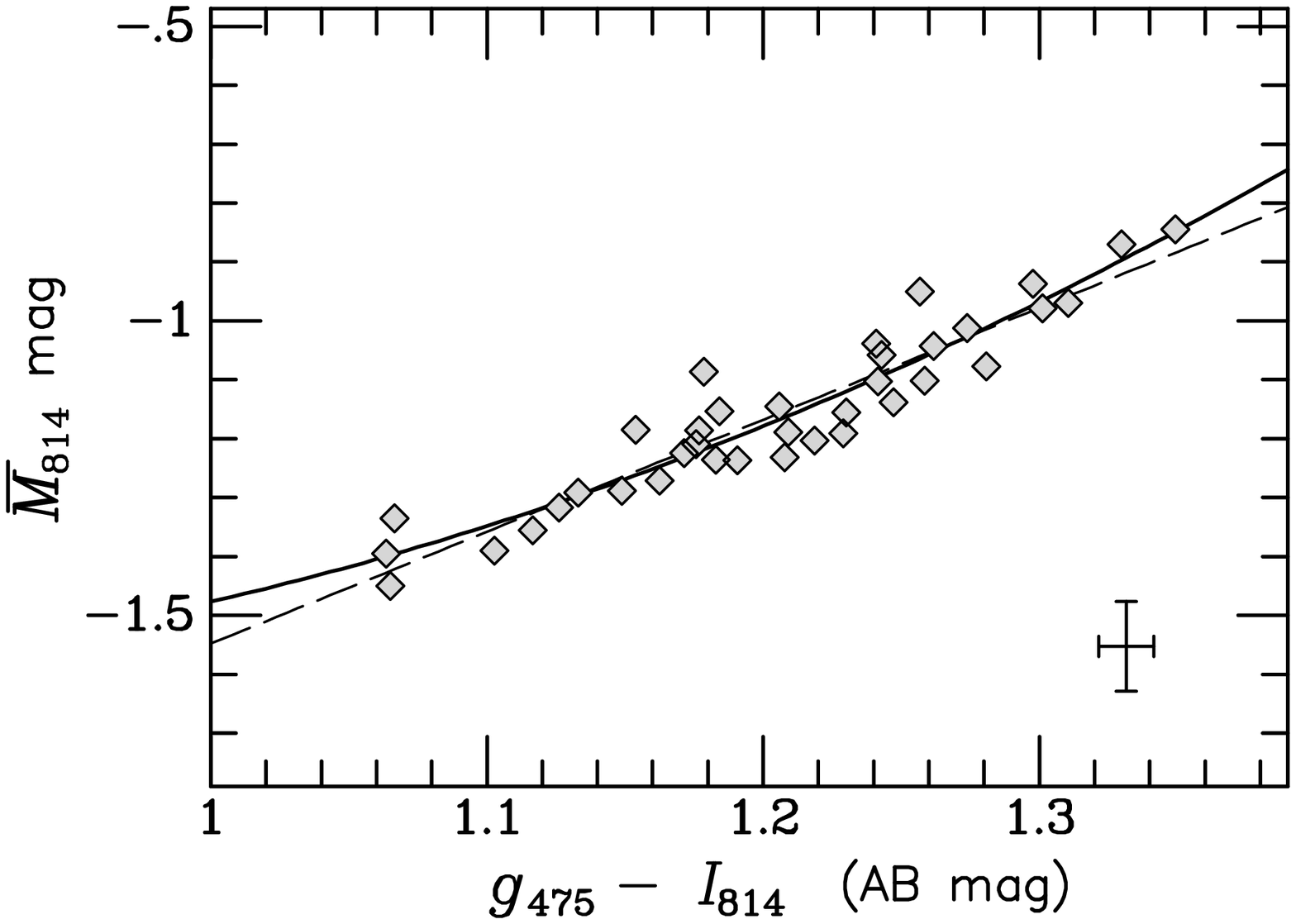}
\caption{%
\textit{Top:}~apparent SBF magnitude as a function of \gIacs\ color for
individual annuli within the nine program galaxies.
\textit{Bottom:}~mean absolute SBF magnitudes after subtracting the
individual distance moduli from \fcsv\ are plotted versus color.
The dashed lines are simple linear least-squares fits, while the solid curves
show quadratic fits for comparison.
Representative measurement errors are indicated in the lower right of each
panel; the \MbarIacs\ error bar includes the typical distance uncertainty.
\label{fig:annuli}}
\end{figure}

The two analysis procedures both measure the \gIacs\ color in each annulus
after applying the same mask as used for the SBF measurement.  We then
calculate the biweight means of the color and SBF measurements from among the
annuli to get the average values for each galaxy.
We note that radial color gradients are common in early-type galaxies and SBF
magnitudes follow the same trends.  SBF and color gradients have been studied
in ACS imaging data from a stellar population perspective by
Cantiello et al.\ (2005, 2007a).  All of the galaxies in our sample
have detectable color gradients except the bluest galaxy NGC\,1375.
However, as found by \fcsv, the presence of such gradients is not a problem
for our averaging procedure as long as the SBF--color relation
is approximately linear over the color range within each galaxy.
Figure~\ref{fig:annuli} shows SBF and color measurements for the
individual annuli within the galaxies.  The top panel plots the observed values
of \mbarIacs, while the lower panel plots absolute \MbarIacs\ after subtracting
the \zacs\ SBF distance moduli from \fcsv.  There is a clear trend between SBF
magnitude and \gIacs\ color, well approximated by a linear relation.
The scatter is about $\sim\,$30\% lower for \MbarIacs\ as compared to
\mbarIacs.  The improvement in scatter for a quadratic fit, as compared to
linear one, is not significant.

The mean difference in galaxy SBF magnitudes from the Tonry and IDL power
spectrum analyses was $0.005\pm0.005$ mag, with an rms scatter of
0.017~mag, which is within the range of the measurement errors. 
Only NGC\,1316 showed an absolute difference as large as 0.03~mag
because of the difficulty in selecting ``clean'' regions in this dusty galaxy;
for the other eight, the difference was 0.02~mag or less.  This represents the
first detailed comparison between the two 
analysis codes that have been used most extensively for SBF measurement.
The close agreement gives us confidence that our SBF
calibration is applicable to either one.  Here we adopt the
results from the IDL analysis simply because it is the same code as
used in our past ACS/WFC studies, and the steps have been
described in detail by those works.  
% However, the results from the Tonry code are nearly identical. 
Table~\ref{tab:tab_result} presents the final SBF and color measurements, as
well as the distance information derived from the calibration presented below.

\subsection{Effect of the Point Spread Function}
\label{ssec:psf}

As noted in the previous section, the SBF measurement involves fitting the
amplitude of the component of the power spectrum that has been convolved with
the PSF of the image.  
Thus, the PSF template is an important consideration in the analysis.
%Thus, the accuracy of the SBF measurement depends on having an accurate PSF template.  
%
The ACSVCS and ACSFCS analyses (Mei \etal\ 2005b; \fcsv) both used a composite
empirical PSF constructed from archival images by Blakeslee \etal\ (2006) for
deconvolving the profiles of high-redshift cluster galaxies.  
The PSF variation within ACS/WFC images after distortion-corrected is small
(Krist 2003) and occurs on small scales (largest wavenumbers) that do not
significantly affect the power spectrum fits (Mei \etal\ 2005a), although a 2\%
error in the fitting for different annuli was assumed in \fcsv\ (as well as
here, since the same code was used) from the spatial variation in the PSF.
However, tests by Mei \etal\ (2005a) and Cantiello \etal\ (2005, 2007b)
using multiple ACS/WFC PSF stars (generally taken from different images)
found variations of 0.02--0.05~mag in the final \mbarIacs\ values.
%
% Extensive testing of multiple PSFs by Blakeslee (1999) on ground-based data
% with significant spatial variations and by Jensen \etal\ (2001) on NICMOS data,
% where the high background can be a problem in characterizing the PSF, found
% rms variations of 0.04--0.06~mag in the final \mbar\ values.
Similar, though slightly larger, variations of 0.04--0.06 mag were found in
tests using multiple PSFs in ground-based data with a variable PSF and in
\hst/NICMOS data (Blakeslee 1999; Jensen \etal\ 2001).
The variations in \mbar\ are due to a combination of effects, including errors in
determining the correct normalization of each PSF template (systematic for a
given data set), and mismatch between the PSF template and galaxy image
(potentially random, if the PSF varies from image to image).

For this study, we searched all of our fields for relatively isolated stars
that were bright but unsaturated, and could be used to make a high
signal-to-noise model of the PSF power spectrum.  The density of stars in these
fields, which are at Galactic latitude $b\approx-54^\circ$, is low.  We found only
one suitable star, which was in the IC\,2006 field, and used this as our
primary PSF template.  We also tried PSFs from other data sets and reran the
full IDL power spectrum analysis to assess the effect on \mbarIacs.  The ACS
Instrument Development Team made available the composite PSFs constructed from the
white dwarf spectrophotometric standard stars analyzed by Sirianni \etal\
(2005) and processed with the same interpolation kernel.  The composite star
for the F814W bandpass has been used for some previous SBF studies (Barber
DeGraff \etal\ 2007; Cantiello \etal\ 2005, 2007a,b).  When we used this
template, our final \mbarIacs\ values became systematically fainter by
0.036~mag.  We also tried individual stars from a random ACS F814W image that
we had available, and the final \mbarIacs\ values became brighter by a similar
amount.  

Consistent with previous studies, these tests indicate variations of
$\sim\,$0.04~mag from the PSF.  At least for the current, fairly homogeneous,
data set, any error of this size resulting from the PSF must be mainly
systematic, affecting all the results in the same way, since the scatter in the
calibration presented below is less than this, and is
consistent with known measurement errors.  The low density of Galactic
foreground stars makes it difficult to characterize the PSFs in the
individual images, but no significant image-to-image variations were found
in the mean full widths at half maximum. The fact that we are using a high
signal-to-noise template from one of our program fields should ensure that the PSF
is accurate for that field, and by extension, given the small scatter that we find
for the final calibration (and the quality of the power spectrum fits), the PSF
template appears accurate for the full sample.  However, it may be beneficial
for future SBF studies that use our calibration to compare the results that
they obtain with the PSF used here (available from the authors).

\begin{figure}\epsscale{1.0}
\plotone{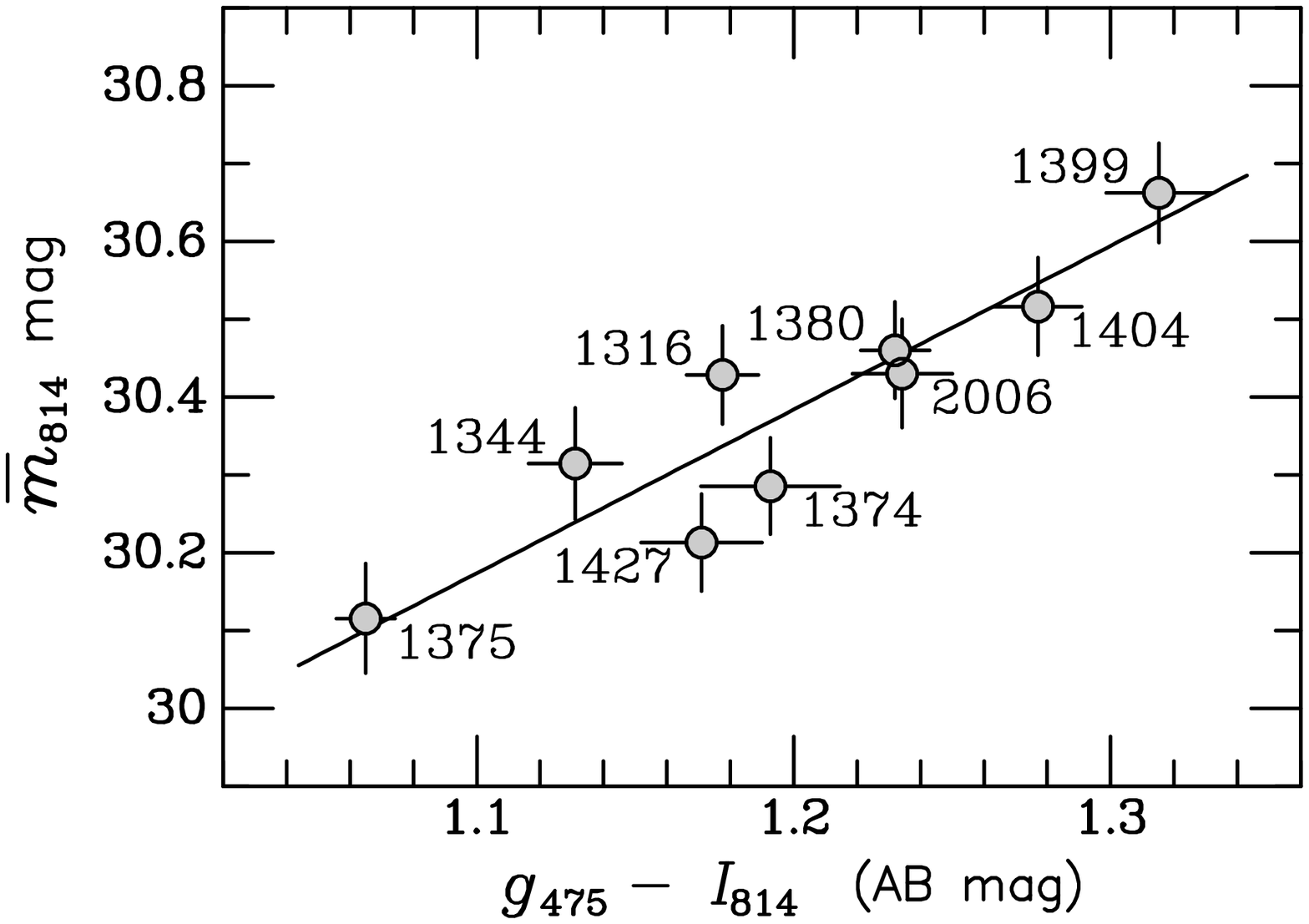}
\vspace{0.5cm}
\plotone{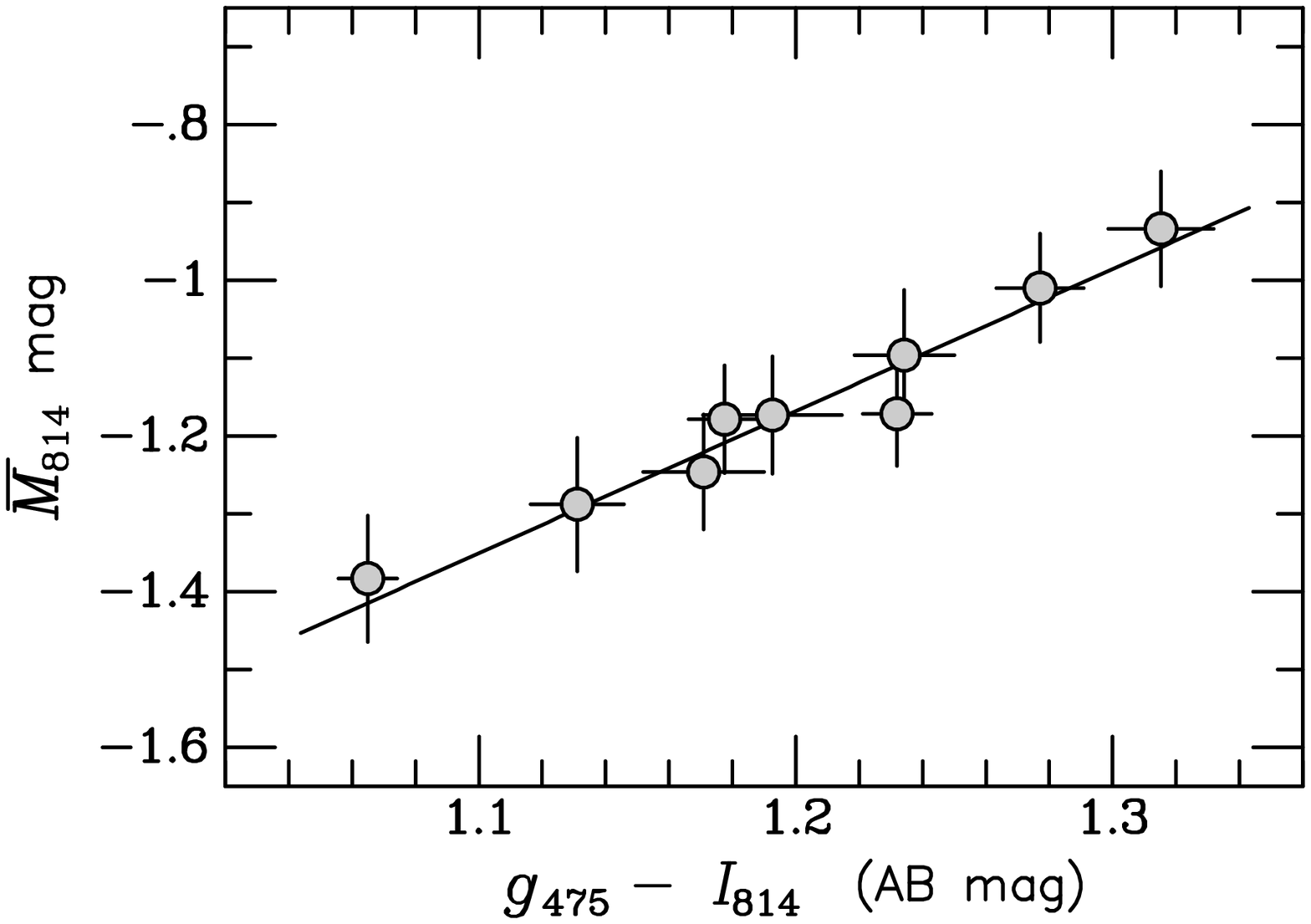}
\caption{\small
The F814W SBF calibration.
\textit{Top:}~mean apparent SBF magnitude as a function of
\gIacs\ color for the sample galaxies.
The vertical error bars are the quadrature sums
of the measurement errors and an intrinsic stellar population scatter of
0.06~mag, as estimated in \fcsv.  Horizontal error bars represent measurement
errors.  Galaxies are labeled by their catalogue numbers.
\textit{Bottom:}~the mean absolute SBF magnitudes after subtracting the
individual distance moduli from \fcsv\ 
are plotted against color.
The vertical error bars are the quadrature sums
of the measurement errors and the distance error (which includes the intrinsic
component) from \fcsv.  Horizontal error bars again represent measurement errors.  
The lines show linear fits including errors in both coordinates; the
coefficients are given in Equations~(\ref{eq:appcal}) and~(\ref{eq:abscal}),
respectively.  
\label{fig:sbcal}}
\end{figure}

\subsection{SBF Calibration for F814W}

% Fig 3
Figure~\ref{fig:sbcal} plots the SBF magnitudes versus \gIacs\ color
for our sample of nine Fornax galaxies, similar to Figure~\ref{fig:annuli}, but
here using the final averaged quantities for each galaxy.
Fitting a linear relation for \mbarIacs\ in terms of \gIacs, 
including errors in both axes and an additional 0.06~mag of
intrinsic (or ``cosmic'') scatter in \mbarIacs\ at fixed \gIacs\ due to stellar
population effects (\fcsv), we~find 
% for the apparent mags:
%  slope = 2.1033 +/- 0.3319  intcpt = 30.3840 +/- 0.0237
%  N =   9   chi2 = 7.00101   prob = 0.428775  rms = 0.06356   chi2/(N-2) = 1
%
\begin{equation}
\mbarIacs \;=\; (30.384\pm0.024) \,+ \,(2.10\pm0.33)\,[\gIacs - 1.2] \,,
\label{eq:appcal}
\end{equation}
where $\gIacs=1.2$ mag is the mean color of our sample.
%, thus minimizing the statistical uncertainty on the zero-point magnitude. 
The quoted error bars reflect the fit uncertainties.  The rms scatter is
0.064~mag, consistent with measurement errors plus the expected 
intrinsic scatter and small depth within Fornax (\fcsv).  

We could shift this calibration by the mean distance modulus of
$31.54\pm0.02$~mag for our nine sample galaxies, but since we have good \zacs\
SBF measurements for each one, and want the calibrations of the two bandpasses
to be as consistent as possible, we instead derive absolute \MbarIacs\
magnitudes using the individual distance measurements.  The lower panel of
Figure~\ref{fig:sbcal} shows the results, which give the following
calibration fit:
% valid for AB colors $1.06<\gIacs<1.32$:
%
%
%  absolute calibration fit again:
%  slope = 1.8260 +/- 0.1866  intcpt = -1.1681 +/- 0.0129
%  N =   9   chi2 = 6.9869   prob = 0.430246  rms = 0.02887   chi2/(N-2) = 0.9981
%     Mbar814 = -1.168 + 1.83[(g-I) - 1.2]
%
% \begin{equation}
\begin{eqnarray}
\MbarIacs &\;=\;& (-1.168\pm0.013\pm0.092)  \nonumber \\
  & ~ & \;+\;  \,(1.83\pm0.20)\,[\gIacs - 1.2] \,,
\label{eq:abscal}
\end{eqnarray}
% {equation}
%
which is valid for AB colors $1.06<\gIacs<1.32$ mag.
The rms scatter of the data points with respect to
this linear calibration is 0.029~mag, and the fit has been derived using the
\gIacs\ and \mbarIacs\ measurement errors, plus 0.015~mag of additional error
added in quadrature to make $\chi^2_n=1.0$ and ensure that the fit errors are
reasonable.  It might seem more correct to include the errors in the individual
distances used in determining the \MbarIacs\ values.  However, when we do this,
we find $\chi^2_n = 0.2$, which has less than a 2\% probability of occurrence.
The likely explanation relates to the behavior of the intrinsic scatter of
0.06~mag that was estimated by \fcsv\ and has been included in the distance errors.
This is the largest component of the error for some galaxies, and apparently it
is strongly coupled between the F814W and F850LP bandpasses, so that \MbarIacs\
and \Mbarzacs\ scatter in the same way (see the following section for more
discussion).  We have therefore omitted this part of the error in deriving the
above fit, and add only 0.015~mag to make $\chi^2_n$ unity.
This 0.015~mag represents an allowance for differential intrinsic
scatter between the two bands, although it may include a small contribution
from some additional measurement error.

The first error bar on the constant coefficient in Equation~(\ref{eq:abscal})
reflects the uncertainty from the fit, which is smaller than for
Equation~(\ref{eq:appcal}) because the scatter of the data points is lower when
using \MbarIacs.  The second error bar gives our estimate of the systematic
uncertainty including the 0.08~mag for the tie of the SBF method to the Cepheid
scale (Tonry \etal\ 2000, 2001; Ferrarese \etal\ 2000), 0.02~mag for the mean
distance of these galaxies from \fcsv, and 0.04~mag for the
possible systematic error in the PSF normalization (Sec.~\ref{ssec:psf}).  It
does not include the additional uncertainty in the Cepheid distance scale,
which may be of order $\sim\,0.1$ mag (Freedman \& Madore~2010).

Table~\ref{tab:tab_result} lists the SBF distance moduli $\mM_{814}$ and
linear distances $d_{814}$ derived for the current sample using the
calibration given by Equation~(\ref{eq:abscal}).  The tabulated distance
uncertainties reflect the total random errors,
including measurement errors and the estimated 0.06~mag of intrinsic
scatter.  They do not include the systematic error in the calibration
zero-point, which is common to all.

\begin{figure}\epsscale{0.95}
\plotone{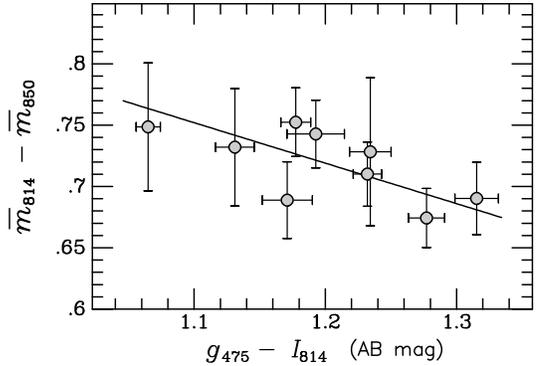}
\caption{%
SBF color $(\mbarIacs{-}\mbarzacs)$ versus integrated color \gIacs\ for the
sample galaxies.
There is some indication that this SBF color becomes slightly bluer as the
galaxies get redder, but the slope of the plotted linear fit is significant at
only the $\lta2\sigma$ level.
\label{fig:sbfcolor}}
\end{figure}

\begin{figure*}\epsscale{0.96}
\plottwo{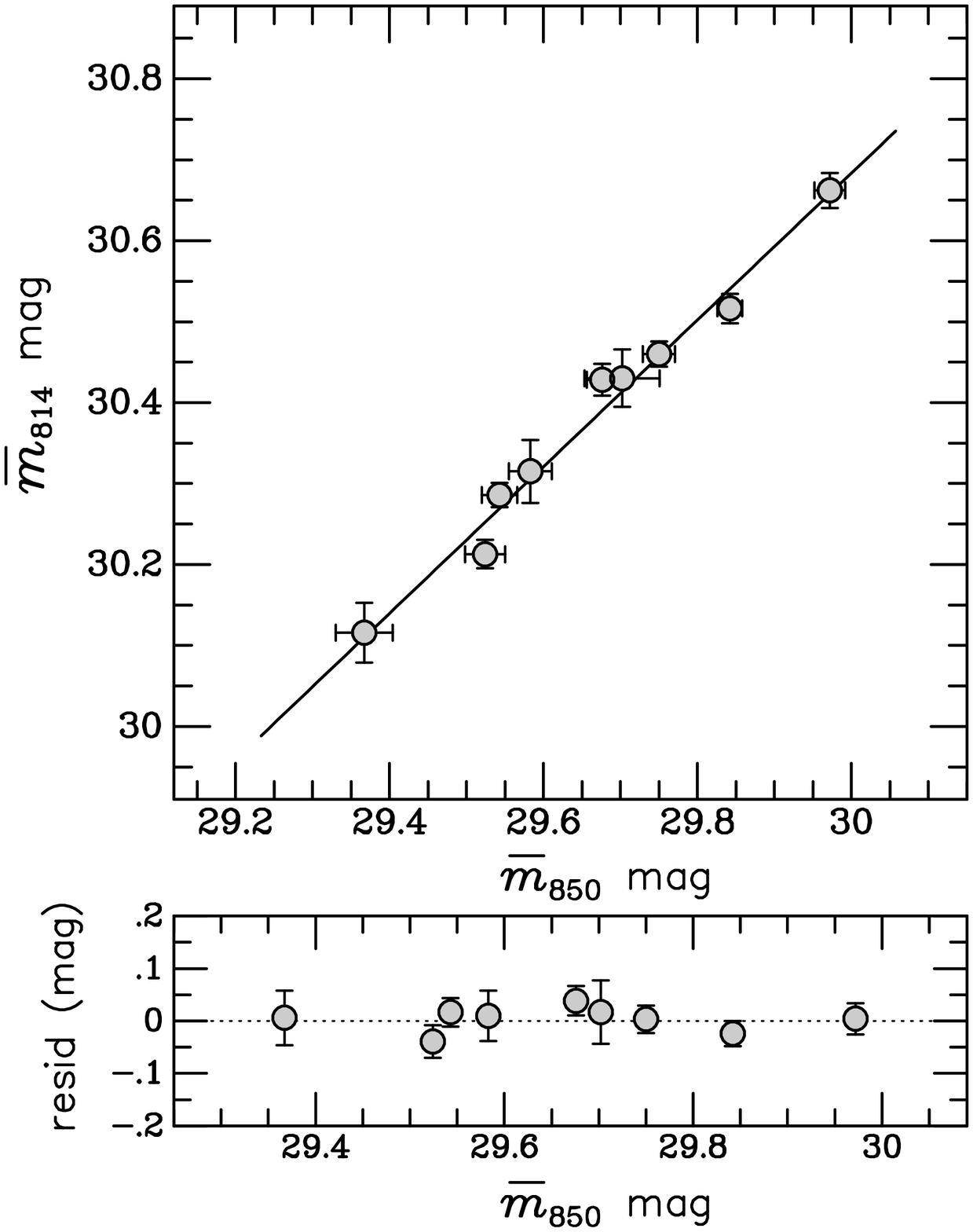}{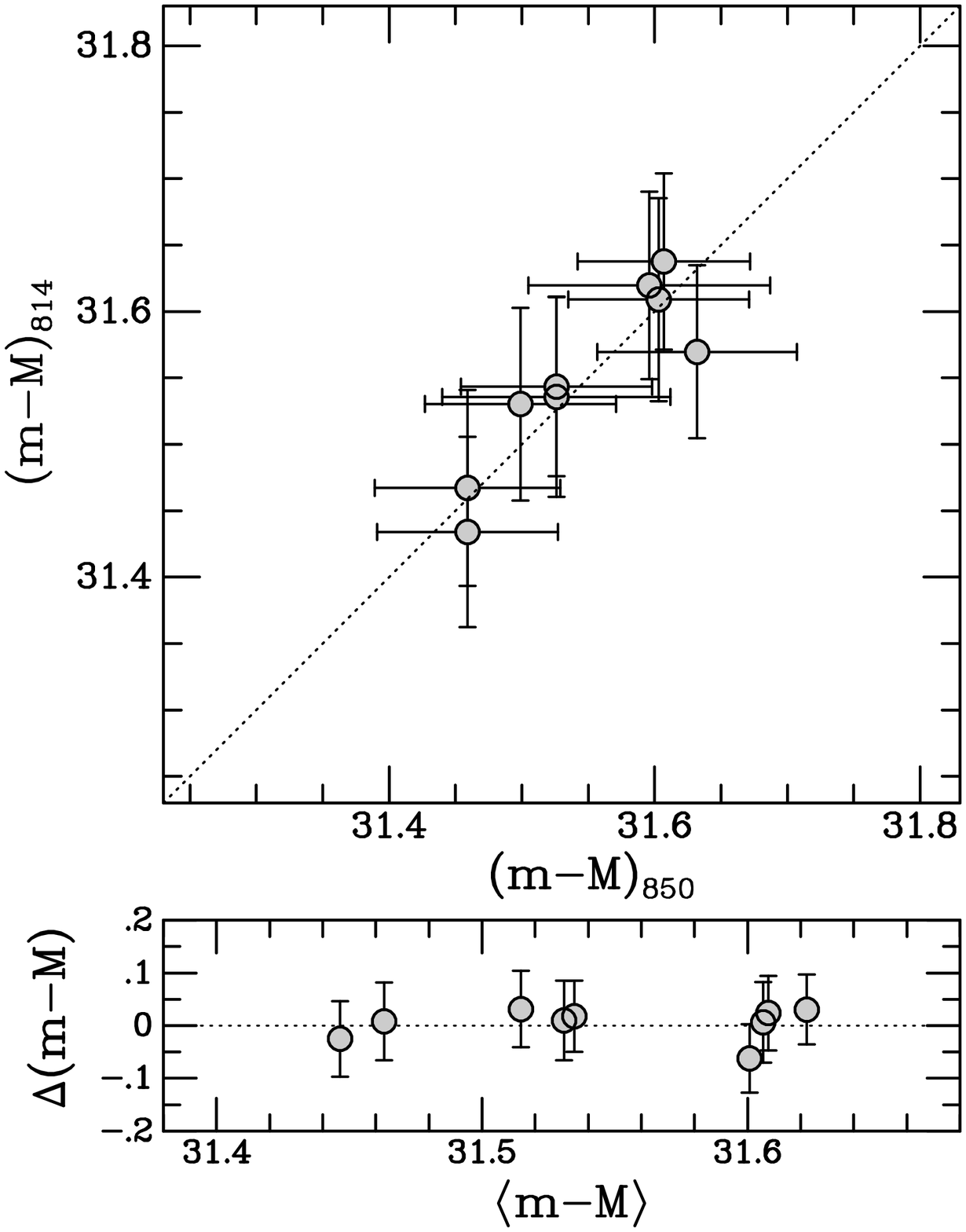}
\caption{%
\textit{Left:}~SBF \mbarIacs\ magnitude from this study is plotted against
\mbarzacs\ from \fcsv.  The errors bars include only measurement error.  The
slope of the linear fit is $0.91\pm0.06$, and the residuals are shown
at bottom, with the error bars here being the quadrature sums of the
measurement errors.
\textit{Right:}~Distance modulus $\mM_{814} \equiv (\mbarIacs{-}\MbarIacs)$ 
from the current study using the
calibration in Equation~(\ref{eq:abscal}) is plotted against distance modulus
$\mM_{850}  \equiv (\mbarzacs{-}\Mbarzacs)$ 
from \fcsv.  The dotted line shows equality, and the differences
between the distance moduli are shown as a function of their average at bottom.  
The intrinsic component of
the errors appears to be strongly correlated between the two bands; see text.
\label{fig:acsfcscomp}}
\end{figure*}

\subsection{Comparison with \zacs-band SBF}

It is useful to compare directly our \Iacs\ SBF magnitudes and distance estimates
with the \zacs\ results from \fcsv.  We find a mean ``SBF color'' for these
galaxies of $\mbarIacs{-}\mbarzacs = 0.72$, with an rms of 0.027~mag.  However,
the SBF color decreases slightly, or gets ``bluer,'' as the SBF magnitudes get
fainter in redder galaxies, as shown in Figure~\ref{fig:sbfcolor}, where
the fitted relation including errors in both coordinates is given by
%
%  slope = -0.3261 +/- 0.1798  intcpt = 0.7186 +/- 0.0111
\begin{equation}
\overline{m}_{814} - \overline{m}_{850} \;=\;
 (0.72\pm0.01) \,- \,(0.33\pm0.18)\,[\gIacs - 1.2] \,,
\label{eq:sbfcolor}
\end{equation}
with an rms scatter of 0.021~mag.  The significance of the slope is
slightly less than $2\sigma$, but the same trend occurs as a function of
\gzacs, with the same scatter.  It is not too unusual for the SBF color to
become bluer while photometric color becomes redder over some range
in population parameters, especially when age is increased at
fixed metallicity (Blakeslee \etal\ 2001).  However, SBF and color
measurements would be needed for more galaxies in more bandpasses to place
strong constraints on the underlying populations.
Put another way, the slope of \mbarIacs\ as a
function of \mbarzacs\ is slightly less than one, as shown in the left panel of
Figure~\ref{fig:acsfcscomp},
% Fig 4
where the plotted linear fit has slope $0.91\pm0.06$. 
% awk 'NR>1 {print $4,$5,$14,$15}' acsfcs5_isbfcal.merge | bivar
%  slope = 0.9064 +/- 0.0616  intcpt = 3.4915 +/- 1.8596
%  N = 9   chi2 = 5.52519   prob = 0.596148  rms = 0.02172   chi2/(N-2) = 0.7893
The rms scatter about this fit is 0.022~mag, which is consistent with
random measurement errors, as the reduced $\chi^2_n = 0.8$ is reasonable for a
linear fit to 9 data points.  This supports our conclusion
above that any intrinsic deviations in \mbarIacs\ and \mbarzacs\ due to
stellar population effects must track closely between the two bands.
% (probability $\sim40$\%).

The right panel of Figure~\ref{fig:acsfcscomp} provides a comparison of the
distance moduli from the present work using Equation~(\ref{eq:abscal}) with those
from \fcsv.  The plotted line indicates unity, and the mean difference between
the two sets of measurements is zero by design, since the 
distance moduli from \fcsv\ went into deriving the
\MbarIacs\ magnitudes used for the \Iacs\ calibration.   The plotted
error bars include the 0.06~mag of intrinsic scatter for each band,
% that is, the random scatter in \mbar\ at a given color.  If 
and naively using these errors, $\chi_n \approx0.1$, 
which has only a $\sim\,$0.2\% probability of occurring.  
Clearly, the intrinsic scatters in the SBF magnitudes in these two
bandpasses are strongly correlated, and the two sets of measurements cannot be
considered completely independent, even though they come from different
data sets in different filters.  However, this comparison is useful in that it
shows the random measurement errors are small as expected, $\lta0.03$~mag.
Further, the tight correlation ensures consistency of the \fcsv\ distances
with those measured here, and thus with future F814W SBF studies
that use our calibration.

\section{Model Comparison and the Distance to Fornax}
\label{sec:models}

The SBF distance zero~point used in this study is an empirical one, ultimately
tied to the metallicity-corrected Cepheid distance scale. 
More specifically, it assumes a mean
distance of 16.5~Mpc to the Virgo cluster based on the Tonry \etal\ (2001)
ground-based SBF distances recalibrated with the final set of \hst\ Key Project
Cepheid distances of Freedman \etal\ (2001); see Mei \etal\ (2005b) and
Appendix~\ref{app:a} of the present work for further details.  The mean distance
of $20$~Mpc for the Fornax cluster then comes from the measured
$0.42\pm0.03$ magnitude offset between the Virgo and Fornax (\fcsv).

There is also a long history of attempts to calibrate the SBF method from
stellar population modeling, thereby making it another ``primary'' distance
indicator not tied to the Cepheids (e.g., Worthey 1993; Buzzoni 1993; Blakeslee \etal\
2001; Mei \etal\ 2001;
Cantiello \etal\ 2003; Raimondo \etal\ 2005; Marin-Franch \& Aparicio
2006).   Biscardi \etal\ (2008) used the Teramo 
``SPoT'' models\footnote{http://www.oao-teramo.inaf.it/SPoT} (Raimondo \etal\ 2005)
in order to derive a theoretical linear calibration of the absolute SBF magnitude 
in the ACS/WFC F814W bandpass as a function of \gIacs\ color.  This is the same
combination of bandpasses as we have calibrated here from an empirical
standpoint, although we work in the AB system while Biscardi used the VEGAMAG
system. The slope is of course unaffected by this difference, and their quoted
value of $2.2\pm0.2$ is close to our empirical results in
Equations~(\ref{eq:appcal}) and~(\ref{eq:abscal}).

\begin{figure}\epsscale{1.15}
\plotone{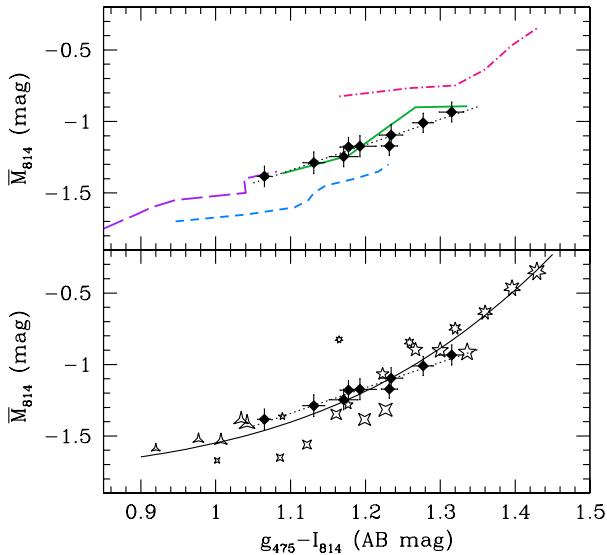}
\caption{%
Comparison of SBF observations with model predictions.
\textit{Top:}~interpolated curves for Teramo
SPoT models with metallicities $[\hbox{Fe/H}] = -0.7, -0.3, 0.0, +0.3$
(long-dashed purple, short-dashed blue, solid green, and dot-dashed red
curves, respectively) 
and ages raging from 3 to 14~Gyr.  The age along each curve generally
increases from left to right with 
photometric color.  The predictions for the solar metallicity models (green solid
curve) lie near the data points (black diamonds); the dotted line shows the
empirical calibration.
\textit{Bottom:}~similar to the top panel, but here the models with different 
$[\hbox{Fe/H}]$ are represented
by symbols having three ($-0.7$), four ($-0.3$), five (solar), and seven
($+0.3$) points.  Symbol size increases with age, and we plot model ages of 
$3,5,7,9,11,$ and 13~Gyr.
The curve shows a cubic polynomial fit to the plotted models.  
Over the range of the observations, the fitted model relation follows the
empirical one fairly well, but the models predict more
curvature beyond this range, similar to what has been observed for \Mbarzacs\
over a larger range of galaxy~colors.
\label{fig:models}}
\end{figure}

Comparisons between empirical calibrations and those derived from simple stellar
population models are complicated because one is never sure which models best
represents actual galaxies.  But it is a useful exercise to check for general
consistency.  Figure~\ref{fig:models} provides a direct comparison of our SBF
and color observations with the predictions from the SPoT models, as in
Biscardi \etal\ (2008) but transformed to
the AB system.  The top panel shows the loci for models with metallicities
$[\hbox{Fe/H}] = -0.7, -0.3, 0.0, +0.3$, and ages ranging from 3 to 14~Gyr for
each.  We also show our measurements from Figure~\ref{fig:sbcal}, along with
the fit from Equation~(\ref{eq:abscal}).
The lower panel is similar, but symbols are used to show the model values for 6
different ages at each metallicity.  The linear empirical calibration overlaps
remarkably well with the locus of the solar metallicity models, which must be at
least partly fortuitous, but consistent with the mean metallicities in these galaxies
being not too far from solar.  The curve in the lower panel is a cubic
polynomial fit to the plotted models.

Two things are particularly worthy of note in Figure~\ref{fig:models}.  First,
there is remarkable consistency between the  zero points of the observations and
models.  In fact, the fits to data and models cross very near the mean color of
$\gIacs = 1.2$.  The excellent agreement between two very different avenues for
calibrating the F814W SBF magnitudes suggests that neither is likely to be
far wrong, and thus the mean Fornax distance of $\sim20$ Mpc appears correct.
Second, the fit to the model predictions indicates some curvature in the relation,
similar to what we found in \fcsv\ for the \mbarzacs\ magnitudes of a larger
sample of galaxies including dwarfs that extended to much bluer colors.
Over the color range of the galaxies in our sample, the deviation
between the linear and cubic fits is small, so the linear calibration
suffices.  However, one should not blindly apply this calibration beyond the
fitted range, as it is reasonable to expect that the curvature may be
significant.

\section{Summary}
\label{sec:summary}

We have measured SBF \mbarIacs\ magnitudes and \gIacs\ colors for nine bright
early-type galaxies in the Fornax cluster from new and archival \hst\ ACS/WFC
imaging data.  We employed two different analysis procedures for measuring the SBF
and found excellent agreement between them.  For this sample of galaxies, which
spans the color range $1.06<\gIacs<1.32$ (AB mag), there is a clear trend of
\mbarIacs\ with \gIacs\ that is well approximated by a linear relation of slope
$\sim2$.  We have used the $\zacs$ SBF distances from \fcsv\ to obtain absolute
\MbarIacs\ for all the galaxies and derive an empirical calibration of the
\Iacs\ SBF method that is consistent with the large sample of existing \zacs\
SBF data.  Because of the greater efficiency of the F814W bandpass, the linear
calibration defined by Equation~(\ref{eq:abscal}) will allow accurate SBF
distances to be measured with the ACS/WFC at greater distances than can be
reached with the \zacs~band in the same amount of time.  
For instance, 4~orbits in \Iacs\ suffices for an
SBF distance to the Coma cluster at 100~Mpc, whereas $\sim\,$10 orbits would be
required in~\zacs.

The observed scatter in the calibration of \MbarIacs\ as a function of \gIacs\
is only 0.03~mag, significantly smaller than the expected $\sim\,$0.06 mag from
intrinsic scatter in the method due to stellar population variations at a given
color (Tonry \etal\ 1997; \fcsv).  
This indicates that there is coherence between the intrinsic scatter in the
\zacs\ and \Iacs\ SBF relations, as expected from the similarity, and even
partial overlap, of these passbands.
Thus, stellar population effects cause the $(\mbarIacs{-}\MbarIacs)$ and
$(\mbarzacs{-}\Mbarzacs)$ distance moduli to scatter in the same direction, so
that they are not fully independent.  Direct
comparison of the two sets of distances supports this conclusion, which helps
to ensure a high degree of consistency between SBF distances measured in these
two bands.

Comparison of our measurements with stellar population model predictions
provides support for the
empirically-based mean distance of 20~Mpc to the Fornax cluster.  According to
these models, and by analogy to the broader-baseline \zacs\ SBF calibration, it
is likely that the \Iacs\ SBF calibration will show curvature beyond the color
limits of our sample.  Therefore, the calibration should not be linearly
extrapolated far beyond the range explored here.  However, our sample was designed
to have colors similar to those of the giant ellipticals targeted in more distant
SBF studies; consequently, there should be little need for extrapolation.  At least two
different SBF studies in the F814W bandpass are ongoing and can benefit from
our calibration.  We have emphasized the need for good PSF templates in SBF
measurements and the importance of investigating systematic variations due to
the PSF.  Our results indicate that when properly calibrated and controlled for
systematics, the SBF method remains one of the most accurate ways of measuring
extragalactic distances.
%
% In the appendices, we provide a simple correction formula for the
% ground-based $I$-band SBF Survey distances and the offset required to make
% published NICMOS SBF distances consistent with later work.
% We also report new \zacs\ SBF distances for two compact elliptical galaxies
% from the ACS Virgo Cluster Survey.

\acknowledgements  
%% this is the official formula (more or less):
Support for this work was provided by NASA through grant number GO-10911 from
the Space Telescope Science Institute, which is operated by AURA, Inc., under
NASA contract NAS~5-26555.  M.C.\ acknowledges support from COFIS ASI-INAF
I/016/07/0 and PRIN-INAF 2008 (PI.\ M.~Marconi).   A.J.\ acknowledges
support from BASAL CATA PFB-06, FONDAP CFA 15010003 and MIDEPLAN ICM
Nucleus P07-021-F.
This research made use of the NASA/IPAC Extragalactic Database (NED) which is
operated by the Jet Propulsion Laboratory, California Institute of Technology,
under contract with NASA.

{\it Facility:} \facility{HST (ACS/WFC)}

% \clearpage
\medskip

\appendix
\section{A. SBF Survey Distance Corrections}
\label{app:a}

\def\qton{\ensuremath{Q_{\rm T01}}}

Tonry \etal\ (2001; hereafter in this appendix, T01) published distances for
300 galaxies from the ground-based $I$-band SBF Survey of Galaxy Distances (Tonry \etal\
1997).  The median quoted uncertainty of 0.22~mag, just over 10\% in distance,
makes this the largest catalogue of nearby galaxy distances with this
level of precision.  It has proven a useful resource for numerous
extragalactic studies.
However, following the publication of our combined sample of 134~galaxies with
high-quality ACS \zacs-band SBF distances (median error 0.075~mag) 
and the direct comparisons for 50 galaxies in common with T01, there has been
some confusion over the possible need for correcting the T01 distances.
As we have received more than a few queries on this issue, we address here
in detail the question of the T01 distances, as well as those of Jensen
\etal~(2003; hereafter J03).

The ground-based SBF survey was conducted over numerous observing runs at
multiple observatories with different types of detectors under variable seeing
and photometric conditions during the course of a decade which saw great
evolution in astronomical instrumentation and our own reduction software.  It
targeted galaxies ranging from the Local Group out to the fuzzy,
condition-dependent limit of the ground-based SBF method.
% perhaps this should be removed...
Although great effort was taken to homogenize the disparate data
sets and produce the best possible set of final distances, the limitations of
the data were already clear at the time; the authors recommended
``further study of the degree of bias present in $[$the T01$]$ data set'' 
(see Sections 4 and 7 of T01).

A separate issue was the zero point.  The calibration of the T01 distance
catalogue was based on comparisons to six galaxies with \hst\ Key Project (KP) Cepheid
distances as tabulated by Ferrarese \etal\ (2000), which yielded an \MbarI\ zero
point of $-1.74$ mag (Tonry \etal\ 2000).  Blakeslee \etal\ (2002) found that
when the calibration was rederived using the revised KP Cepheid distances
from Freedman \etal\ (2001), the resulting zero-point shifted fainter to
$-1.68$~mag.  If there were no distance-dependent biases, then this simple shift in zero
point would mean that all the T01 distance moduli just needed to be revised
downward by 0.06 mag.  For instance, our ACSVCS papers applied this shift
to the mean Virgo distance presented by T01, thus
adopting 31.09 mag, or 16.5 Mpc, for Virgo.  This was a reasonably ``safe''
assumption, given that the six galaxies used to tie together the SBF and Cepheid
distances extended out to the Virgo cluster, and were generally targeted in
above-average observing conditions.  Note that some researchers have
misinterpreted this 0.06 mag shift as resulting from a change in the
adopted Hubble constant, but there is no explicit dependence on
the Hubble constant, only a dependence on the zero point of the Cepheid
distances.

In addition to rederiving the distance zero point based on the revised KP Cepheid
distances, Blakeslee \etal\ (2002) found some evidence for distance bias in the
lowest quality T01 data, in the sense that the relative distances 
tended to be underestimated for the lower quality data, as suspected by T01.
This conclusion was based on comparisons with distances estimated from the
fundamental plane and density field-corrected velocity measurements.
Note that the absolute data quality generally decreases with distance, e.g., the
errors increase with distance in T01, so this could potentially translate into a
distance-dependent bias.  Because of this combination of zero
point and bias issues,  Blakeslee \etal\ (2009) chose to 
compare the distances for the 50 galaxies in common between the T01 and ACS 
data sets without first applying any shifts to T01.  This contrasted with our
approach in Mei \etal\ (2007), where the comparison was made only for Virgo
galaxies and the zero-point shift from Blakeslee \etal\ (2002) was applied; this
difference may have been the cause of some subsequent confusion.
Blakeslee \etal\ (2009) found the same mean Fornax distance modulus to within
$\pm0.01$ mag as T01, without any shifting in zero point.  However, the T01 Virgo
distance modulus is then necessarily too high by $0.06$ mag.
As compared to the ACS
studies, the data tabulated by T01 therefore give a smaller relative
distance modulus of Fornax with respect to Virgo, although the uncertainties overlap.
In light of these results, it becomes unclear whether or not a simple shift 
in the zero point of the T01 distance moduli is useful without a second-order
correction that may depend
on the distances of the galaxies under consideration.  Given the
0.22~mag median error of the T01 distances, 
it could be argued that these effects are in the noise for
any given galaxy.  For galaxies within Virgo and Fornax, the tabulated ACS distances
from Blakeslee \etal\ (2009) are much more accurate, and negligibly affected by
the background variance corrections that were suggested to bias the ground-based data.

However, for some purposes, it may be desirable to have a general
correction formula for the published T01 distances, including both the zero point
and second-order bias corrections.
Given our high-quality measurements for a 
sizable fraction of the T01 sample, we have attempted to derive one.
We investigated possible correlations of the distance modulus difference
$\Delta_{\rm T01} \equiv \mM_{\rm ACS} - \mM_{\rm T01}$, 
where $\mM_{\rm ACS}$ comes from Blakeslee \etal\ (2009),
with various quantities tabulated by T01 so that the correction is available
for that full sample.  In general, the correlations are weak,
except between $\Delta_{\rm T01}$  and $\mM_{\rm T01}$; 
there is no significant correlation with $\mM_{\rm ACS}$. 
This is easily understood if the ACSFCS-V distances are taken
as the ``true'' values, or at least unbiased and homogeneous in quality; then
the distance measurement error $\Delta_{\rm T01}$ will correlate strongly with
the measured distance over any limited range in true distance.  But this
correlation, caused by measurement error, cannot be used to bias-correct the
measured distances for the full sample.  There is no correlation of 
$\Delta_{\rm T01}$ with the quoted measurement error, perhaps because of the
relatively small range in the latter quantity and the multitude of effects that
contribute to it.  

\begin{figure}\epsscale{0.6}
\bigskip
\plotone{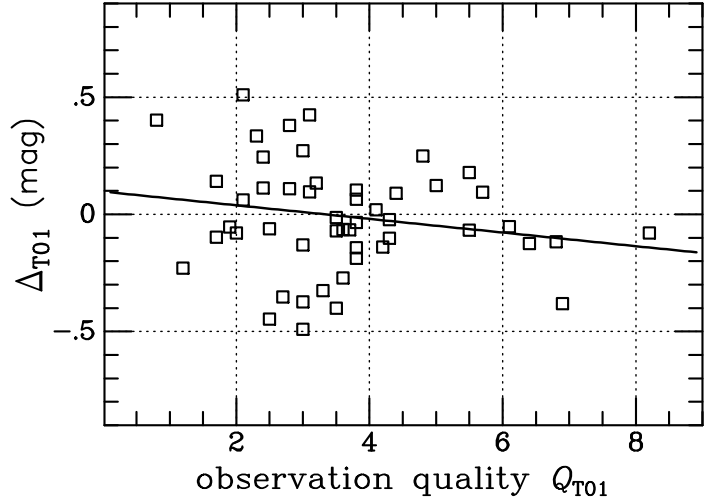}
\caption{%
Difference between the Tonry \etal\ (2001) distance moduli and the \fcsv\
distance moduli, defined such that
 $\Delta_{\rm T01} = \mM_{\rm ACS} - \mM_{\rm T01}$,
is plotted
as a function of the Tonry \etal\ observation quality~\qton\ for the 50 galaxies in
common between the two samples.  Note that the typical error on
$\Delta_{\rm T01}$ is $\sim0.23$ mag, so the scatter is expected to be
large.  See Appendix for details.
\label{fig:qton}}
\smallskip
\end{figure}

It seemed more promising to investigate correlations with parameters that
directly reflect the data quality.  It is also necessary for the correction parameter 
to span reasonably well the same range in the current subsample as in the complete
T01 catalogue.  T01 supplied two measures of the quality of the data for each galaxy.
One such parameter, called $PD$, was the normalized product of the width of
the seeing disk in arcseconds and the expected distance based on the galaxy
recessional velocity in units of 1000 \kms, thus making it independent of
any potential distance biases in the SBF distance measurement itself.  $PD$
is proportional to the linear size of the resolution element at the distance
of the galaxy, and thus is smaller for better quality data.  Another
measure of data quality, called $Q$ by T01 and here referred to as \qton,
scaled logarithmically with the ratio of the signal-to-noise and $PD$; thus
it is larger for better quality data. 
Blakeslee \etal\ (2002) investigated bias as a function of $PD$; we have found a
slightly more significant trend as a function of \qton.
% Fig 6
Figure~\ref{fig:qton} shows $\Delta_{\rm T01}$ plotted against \qton. 
The fitted relation suggests that the 
T01 distance moduli can be bias-corrected according to
%
%\begin{equation}
\begin{eqnarray}
\label{eq:ton01cor}
(m{-}M)_{\rm T01,\,cor} &=& (m{-}M)_{\rm T01,\,raw} \,+\, 0.1 \,-\, 0.03\,\qton\\ 
                        &=& (m{-}M)_{\rm T01,\,raw} \,-\, 0.03(\qton - 3.3)
                        \,. \nonumber
\end{eqnarray}
% \end{equation}
%
Thus, as published without any zero-point shifting, 
the poorest quality T01 distances with $Q_{\rm T01}\approx1$--2 would be
corrected by $\sim\,+$0.05 mag, and the best quality ones with
$Q_{\rm T01}\approx7$--8 would be corrected by $\sim\,-$0.12 mag.  Given
the range of \qton\ in Figure~\ref{fig:qton}, we do
not recommend applying corrections any larger than this.
The six Cepheid calibrator galaxies have an average
$\langle\qton\rangle = 6.0$, giving $\langle\Delta_{\rm T01}\rangle = -0.08$~mag,
(or, omitting the Local Group galaxy M31: $\langle\qton\rangle = 5.2$, 
$\langle\Delta_{\rm T01}\rangle = -0.06$ mag), very close to the expected shift
from the Cepheid calibration, which is included within the correction formula.
% The average correction is roughly similar to the uniform shift discussed above.
%
The significance of the fitted slope is only $1.75\sigma$ based on these 50
galaxies, and the relative Fornax--Virgo distance modulus from the
sample is unchanged,
but if the trend persisted for the full T01 sample of 300 galaxies, the
significance 
would reach $4\sigma$; therefore studies relying on many T01 galaxy distances
may benefit from applying a correction such as Equation~(\ref{eq:ton01cor}).

Finally, we address the J03 \hst/NICMOS F160W SBF distances for 65 galaxies.
These were tied to the KP Cepheid distances from Freedman \etal\ (2001), but
without the espoused $-$0.2 mag~dex$^{-1}$ metallicity correction.
The evidence for this correction was ambiguous at the time, and J03
% made extensive comparisons of their distance measurements with the SBF
% predictions of several sets of stellar population models, finding that the
found that the agreement with SBF predictions from stellar population models
was better when the distance zero point was based on the Cepheid
results without metallicity correction.  Since then, there has been additional
evidence to support a Cepheid metallicity dependence (Sakai \etal\ 2004; Macri
\etal\ 2006; Scowcroft \etal\ 2009), and better understanding of the
uncertainty in model predictions for near-IR SBF
(Raimondo 2009; Gonz{\'a}lez-L{\'o}pezlira 2010; Lee \etal\ 2010).  Given the
high signal-to-noise and good resolution of the J03 NICMOS data, there is no
reason to suspect a bias similar to that of T01.  However, 
% to bring the distances into agreement with those presented here,
% it is necessary to shift
to bring them into agreement with other work requires putting
them on the Cepheid scale that includes the metallicity correction.
J03 derived their distance zero point by comparison to
47 T01 galaxies after shifting the distance moduli 
by $-0.16$~mag, which comes from repeating the Tonry  \etal\
(2000) calibration using the revised Cepheid distances
without metallicity correction.  As noted above, the change in the T01
moduli when calibrated on the revised Cepheid distances with metallicity
correction is $-0.06$ mag.  This suggests a shift of $+0.10$ mag to the Jensen
\etal\ distance moduli, although there is some question about possible
bias in the particular subsample of T01 that was used.  J03 also measured
SBF in nine KP Cepheid galaxies and tabulate the resulting
$\Mbar_{160}$ values for the distances with and without metallicity
correction; the mean difference is 0.08~mag,
% (whether or the sample is restricted to those that J03 consider ``dust free''),
% , and if we restrict this to just the three galaxies that J03 consider
% ``dust free,'' then the mean difference for those is also 0.08~mag, 
which is consistent with the shift in the T01
calibration caused by the metallicity correction.
We therefore conclude that the J03 distance moduli should be increased uniformly by
$+$0.1~mag to bring them into consistency with the corrected T01 distances, our
recent ACS $\zacs$ SBF measurements, and the present work.
%
%

%\begin{figure}\epsscale{0.73}

\section{B. Additional ACSVCS Dwarf Galaxy Distances}
\label{app:b}

Mei \etal\ (2007) published SBF measurements for 90 galaxies from the ACSVCS
\gacs\ and \zacs\ imaging data, and Blakeslee \etal\ (2009) tabulated
recalibrated distances for these along with the 43 ACSFCS galaxies and
NGC\,4697 in the Virgo Southern Extension.  No SBF measurements were published
for ten of the ACVCS galaxies, mainly owing to irregular morphologies.
However, the problem for the galaxies VCC\,1192 and VCC\,1199 was different:
they are both compact elliptical companions of the Virgo brightest cluster
galaxy M49 (NGC\,4472 or VCC\,1226) and the strong background gradient from the
giant galaxy's halo caused the photometric measurements for these galaxies to
be unreliable.  Since the physical distances from M49 of these likely stripped 
(Ferrarese \etal\ 2006; \cote\ \etal\ 2010) companions are of
interest, we have undertaken new measurements of these two galaxies.
Although not directly related to the main content of this paper, these data
are part of our larger \hst\ SBF effort, and we present them here.

We constructed initial isophotal models of the light distributions of 
VCC1192 and VCC1199 in each filter.  We then subtracted these models, masked
the background sources, and interpolated two-dimensional cubic spline surfaces
on a $16{\times}16$ grid (each grid cell being $\sim270$ pix on a side) to make
smooth fits of the background light distribution.  Since these galaxies
are so compact, and they had been removed by the initial models, they
did not perceptibly affect the fits.  After subtraction of the fitted surfaces,
the backgrounds were extremely flat.  We then constructed a new galaxy model
for each
and proceeded with the SBF and color measurements as usual, using the same code
as in the present work.  Table~\ref{tab:vcc119x} presents our measurements for
these galaxies in the same format as for the other ACSVCS galaxies in Table~2
of Blakeslee \etal\ (2009).  This brings the total number of galaxies with
measured \zacs\ SBF distances to 136.

\newpage

\clearpage

% table generated by textab.py !
\begin{deluxetable}{lrcrccccc}
\tabletypesize{\small}
\tablecaption{Galaxy Data}
 \tablewidth{0pt}
\tablehead{
\colhead{Galaxy} &
\colhead{$B_T$} &
\colhead{Morph} &
\colhead{$v_h$} &
\colhead{Other~} &
\colhead{$T_{\rm F475}$} &
\colhead{$T_{\rm F814}$} &
\colhead{$\mM_{850}$} &
\colhead{Program}
 \\
\colhead{(1)} &
\colhead{(2)} &
\colhead{(3)} &
\colhead{(4)} &
\colhead{(5)} &
\colhead{(6)} &
\colhead{(7)} &
\colhead{(8)} &
\colhead{(9)}
}
\startdata
NGC\,1316 & 9.4 & S0\,pec& 1760 & FCC\,21   & 760 & 4680  & $31.607\pm0.065$ & 10217,9409 \\ 
NGC\,1344 & 11.3 & E5    & 1169 & NGC\,1340 & 760 &  960  & $31.603\pm0.068$ & 10217,9399 \\ 
NGC\,1374 & 11.9 & E0    & 1294 & FCC\,147  & 680 & 1224  & $31.459\pm0.070$ & 10911 \\ 
NGC\,1375 & 13.6 & S0    &  740 & FCC\,148  & 680 & 1224  & $31.499\pm0.072$ & 10911 \\ 
NGC\,1380 & 11.3 & S0/a  & 1877 & FCC\,167  & 680 & 1224  & $31.632\pm0.075$ & 10911 \\ 
NGC\,1399 & 10.6 & E1    & 1425 & FCC\,213  & 680 & 1224  & $31.596\pm0.091$ & 10911 \\ 
NGC\,1404 & 10.9 & E2    & 1947 & FCC\,219  & 680 & 1224  & $31.526\pm0.072$ & 10911 \\ 
NGC\,1427 & 11.8 & E4    & 1388 & FCC\,276  & 680 & 1224  & $31.459\pm0.068$ & 10911 \\ 
IC\,2006 & 12.2  & E2    & 1382 & ESO\,359-07& 680 & 1224 & $31.525\pm0.086$ & 10911 \\ 
\enddata
\label{tab:tab_obs}
\tablecomments{Columns list:  (1)~galaxy name;
  (2)~total $B$~magnitude from the FCC (Ferguson 1989) or NED;
  (3)~morphological type from the FCC or NED;
  (4)~heliocentric velocity from NED (\kms)
  (5)~alternative galaxy designation; 
  (6)~\hst/ACS exposure time for the F475W bandpass (sec);
  (7)~\hst/ACS exposure time for the F814W bandpass (sec);
  (8)~distance modulus and uncertainty from \zacs-band SBF (ACSFCS-V);
  (9)~\hst\ program IDs for the observations used in this study.
}
\end{deluxetable}

% table generated by textab.py !
\begin{deluxetable}{lcccc}
\tabletypesize{\small}
\tablecaption{Galaxy Color and SBF Measurements}
\tablewidth{0pt}
\tablehead{
\colhead{Galaxy} &
\colhead{\gI} &
\colhead{$\overline{m}_{814}$} &
\colhead{$\mM_{814}$} &
\colhead{$d_{814}$~(Mpc)}
 \\
\colhead{(1)} &
\colhead{(2)} &
\colhead{(3)} &
\colhead{(4)} &
\colhead{(5)} 
}
\startdata
NGC\,1316 & $1.177\pm0.011$ & $30.428\pm0.020$ & $31.638\pm0.066$ & $21.3\pm0.7$ \\ 
NGC\,1344 & $1.131\pm0.015$ & $30.315\pm0.039$ & $31.609\pm0.076$ & $21.0\pm0.7$ \\ 
NGC\,1374 & $1.193\pm0.022$ & $30.286\pm0.015$ & $31.467\pm0.074$ & $19.7\pm0.7$ \\ 
NGC\,1375 & $1.065\pm0.009$ & $30.116\pm0.037$ & $31.530\pm0.072$ & $20.2\pm0.7$ \\ 
NGC\,1380 & $1.232\pm0.011$ & $30.460\pm0.016$ & $31.570\pm0.065$ & $20.6\pm0.6$ \\ 
NGC\,1399 & $1.315\pm0.017$ & $30.662\pm0.022$ & $31.620\pm0.071$ & $21.1\pm0.7$ \\ 
NGC\,1404 & $1.277\pm0.014$ & $30.516\pm0.018$ & $31.544\pm0.068$ & $20.4\pm0.6$ \\ 
NGC\,1427 & $1.171\pm0.019$ & $30.213\pm0.017$ & $31.434\pm0.072$ & $19.4\pm0.6$ \\ 
IC\,2006 & $1.234\pm0.016$ & $30.430\pm0.035$ & $31.536\pm0.075$ & $20.3\pm0.7$ \\ 
\enddata
\label{tab:tab_result}
\tablecomments{Columns list:  (1)~galaxy name;
  (2)~mean galaxy \gIacs\ color;
  (3)~mean SBF magnitude \mbarIacs;
  (4)~distance modulus derived from the linear \Iacs\ SBF calibration presented here;
  (5)~the distance in Mpc.
The distance errors include contributions from SBF and color measurement
errors, as well as the estimated intrinsic (``cosmic'') scatter in the method.}
\end{deluxetable}

% put in format for the table...
% table generated by textab.py !
\begin{deluxetable}{lrrrrrc}
\tabletypesize{\small}
\tablecaption{Distances for Additional ACS Virgo Cluster Survey Galaxies}
\tablewidth{0pt}
\tablehead{
\colhead{Galaxy} &
\colhead{$(g{-}z)$} &
\colhead{$\mbarz$} &
\colhead{\mM} &
\colhead{$d$} &
\colhead{$B_T$} &
\colhead{Name} \\
\colhead{(1)} &
\colhead{(2)} &
\colhead{(3)} &
\colhead{(4)} &
\colhead{(5)} &
\colhead{(6)} &
\colhead{(7)}
}
\startdata
VCC1192 & $1.443\pm0.009$ & $29.305\pm0.065$ & $31.078\pm0.091$ & $16.4\pm0.7$ & 15.0 & NGC4467 \\ 
VCC1199 & $1.469\pm0.030$ & $29.345\pm0.169$ & $31.055\pm0.196$ & $16.3\pm1.5$ & 15.5 & \dots \\ 
\enddata
\tablecomments{Columns list:  (1) VCC designation (Binggeli \etal\ 1985); 
  (2)~mean \gz\ color of the analyzed region;
  (3)~mean F850LP SBF magnitude \mbarz;
  (4)~distance modulus derived from Equation\,$[6]$ of \fcsv; 
  (5)~distance in Mpc; 
  (6)~total $B$~magnitude from the VCC or NED; 
  (7)~common galaxy name. 
}
\label{tab:vcc119x}
\end{deluxetable}

\end{document}